# *On the azimuthal dependence of orthorhombic fractures for vertical seismic profiles using the ellipsoidal approximation*

P. Contreras *
Physics Department, University of Los Andes, Mérida, 5001, Venezuela.
* Corresponding author (pcontreras@mailfence.com)

## ABSTRACT

The analysis of travel times with move-out velocities represents one of the most widely used seismic signal processing techniques for the exploration and monitoring of Oil and Gas reservoirs. In travel time analysis for anisotropic elastic media with fractures, the knowledge of the conversion point at interfaces with incident longitudinal, and reflected transversal elastic response impulses that take into account the azimuthal dependence is important for seismic data analysis acquired in multicomponent surveys, and for the correct interpretation of seismic in reservoirs. Henceforth, this work shows how to derive analytical expressions for the longitudinal to transverse response impulse" P-$S_i$" conversion points using the ellipsoidal approximation for fractured orthorhombic elastic media under kinematical considerations. These expressions can be used for vertical seismic profiles with small polar angles of aperture and azimuthal dependence in orthorhombic media when the elastic slowness is used as a main theoretical tool to solve the Christoffel equation. We also explain some of the differences within the ellipsoidal inversion procedure of the elastic stiffnesses $c_{13}$ and $c_{23}$.

**Keywords:** Vertical Seismic Profiles, Ellipsoidal Approximation, Orthorhombic Anisotropy, *P-$S_i$* Conversion Points, Azimuthally Dependent Elastic Media.

## INTRODUCTION

In seismic Oil and Gas exploration, one of the most important theoretical tools to start with any analysis that involves the direct task of finding the elastic velocity field to infer lithology properties, the inverse estimation of the elastic stiffness tensor field and the fracture orientation is the Christoffel equation [1,2] and references therein. The Christoffel equation can be solved to initially give the eigenvalues and eigenvectors that represent phase velocities and phase angles for all elastic symmetries [2] (the symmetries of anisotropic rocks in seismic are a subgroup of the symmetries found in the Bravais lattice, namely, the hexagonal (VTI), the orthorhombic, the monoclinic and the triclinic ones. They are called low symmetry systems and with the subsequent estimation and visualization of the group velocity elastic field (also called response impulses field) and their corresponding angles calculation, (see Fig. 1 panel a) [3] serve as the main tool for what is called in Exploration Geophysics the forward modeling of the Earth's crust. This procedure finds practical applications in VTI media with a vertical axis of symmetry [3], and also in azimuthally dependent [4] fracture media such as the HTI [5], the orthorhombic [6], and the monoclinic elastic fractured systems [7]. However, there are additional issues on how to apply the elasticity formalism derived from the Christoffel equation to seismic exploration.

The first factor to consider is the acquisition geometry if the acquisition has wider angles of apertures with a surface source and geophones arrangement, or on the other hand, vertical seismic profiles (VSP) with a surface source and downhole geophones. This is a practical issue since it takes into consideration the data acquisition and processing in geophysical exploration and monitoring analysis of Oil wells. It can in principle be used sources with P waves and additionally use the transversal waves generated at the interface when the elastic energy converts from one to another elastic mode (P – $S_1$ and P – $S_2$ conversion points). Additionally, it can be used multicomponent seismic records with 3D sources and 3D geophones able to generate and record the transversal elastic field (multicomponent seismic). If the whole elastic field is included in the source and it is performed in multicomponent surveys and in addition to petrophysics analysis considering vertical seismic profiles and Crosswell geometries, the quantity of information to be processed increases considerable and any theoretical tool available helps in the task of imaging the subsurface with smaller error percentage.

Historically, most of the seismic signal anisotropic processing analysis done has been based on the Thomsen parametrization [3] or the derivations from it that consider azimuthally dependence of the wave field [4]. Henceforth, the second point to consider is theoretical, and addresses which variables and approximations can be used to solve the Christoffel equation [1]. This depends on the first factor since the acquisition of seismic data depends on what approximation is used to solve the velocity elastic field or the inverse estimation of the elastic stiffnesses.

On the other hand, the solution of the Christoffel equation in Oil and Gas shale reservoirs, despite the enormous advance during all these years and which started with the use of the Thomsen analytical approximation for VTI systems [3], has evolved over time. One can use the classical solution for the Christoffel equation [2] and compute phase variables (velocities and angles) and find group variables (group velocities and angles). That can be done for both acquisition geometries, surface-to-surface, and VSP. The slowness solution is also suitable and serves for both types of acquisition geometries including directly non-hyperbolic move-out analysis [4,8]. These are two common approaches used in the exploration of geophysics for Oil and Gas purposes. A new theoretical tool was proposed recently, the Christoffel equation in the polarization variables [9] which is important to mention for future practical applications. Therefore, we have introduced Table 1 with a brief classification of the main theoretical methods to solve the Christoffel equation and some of the first references to appear in the literature:

*Table 1: Different solutions for the eigenvalues of the Christoffel equation using the normal wave front, the slowness and the polarization vector.*

| Eigenvalues of the Christoffel equation for phase velocities $\Gamma(\mathbf{n})$, as function of the wave front normal vector $\mathbf{n}$. These eigenvalues can be used for both types of acquisition geometries [1,2,3]. | Eigenvalues of the Christoffel equation $\Gamma(\mathbf{p})$ for phase slowness $\mathbf{p}$. These eigenvalues can be used for both kinds of acquisition geometries [4,8]. | Eigenvalues of the Christoffel matrix $\Gamma(\mathbf{U})$ for the polarization vector U, when they are considered as variables of the phase slowness function $\mathbf{p(U)}$ [9]. |
|---|---|---|

To summarize, rocks can be anisotropic for various reasons, such as the presence of fractures or strong lithological changes [3,4,6,8]. Omitting the presence of anisotropy can cause distortions in seismic imaging that turns into economic losses due to well-location errors. On the other hand, Anisotropy can be in some cases the answer to the problems of identifying different types of lithology. Therefore, the characterization of anisotropy by estimating elastic constants is essential for the development of velocity estimation techniques, modeling, and 2D and 3D seismic processing in order to reduce exploration risk.

In this work, we outline the theoretical computational work performed using the so-called ellipsoidal orthorhombic approximation that was introduced years ago [15,16]. We show how to use their azimuthal dependence and also the way they are visualized in the symmetry vertical planes of orthorhombic media with their main implications as one more approximation to the complex field of elastic wave propagation in low symmetry systems, and finally, we give a theoretical conversion P – Si point derivation for the ellipsoidal orthorhombic case that can be used in anisotropic signal processing.

**ELLIPSOIDAL SEISMIC VELOCITIES IN ORTHORROMBIC MEDIA FOR VERTICAL SEISMIC PROFILES.**

The analysis using the so-called normal move-out velocities was started years ago; it proposed a wave equation based on normal move-out velocities [10]. The use of elliptical seismic velocities was widely used for isotropic and anisotropic media [11,12,13,14]. Furthermore, in the work [15] it was shown that using a Taylor expansion near the vertical planes of symmetry for an orthorhombic elastic media, one could approximate the whole elastic field by obtaining theoretical expressions for the phase and group seismic velocities and angles near the two vertical axes using Silicon graphic stations for their visualization. Sometime later, the calculation of the response impulses was validated using an inversion procedure [16] to obtain the elastic constants for the Cracked Greenhorn Shale [14] and was considered of practical interest to test vertical seismic profiles acquisitions in orthorhombic media. That procedure was performed with the help of the eigenvalues of the Christoffel equation $\Gamma(\mathbf{n})$ using the wave front normal vectors n with azimuthal dependence following [2] and the idea between phase and group velocities developed in [13] compared with the classical approach where the eigenvalues determinant $F = \text{Det}\,(\Gamma_{ik}(\mathbf{n}) - \rho\, v^2\, \delta_{ik}) = 0$ allows to solve for phase velocities [2].

Sometime later, V. Grechka and I. Tsvankin pointed out to the author of this work, that the same procedure could be obtained using the eigenvalues of the Christoffel matrix $\Gamma(\mathbf{p})$ for the phase slowness p, with several advantages [7]. Among them, the eigenvalues of the slowness vector p can be used for both kinds of acquisition geometries by means of the Taylor expansion that allows hyperbolic or non-hyperbolic normal move-out analysis using the second or the fourth derivative terms where the azimuthal variables are explicit functions of the vertical slowness q of Eq. 1. Henceforth, following the general expansion for low symmetry elastic systems with azimuthal dependence in terms of the slowness, the general equation is written as a Taylor expansion [7]

$$q(p_x, p_y) \simeq q^0 + q_i^0 \, p_i + \frac{1}{2} \, q_{ij}^0 \, p_i \, p_j + \frac{1}{6} \, q_{ijk}^0 \, p_i \, p_j \, p_k + \frac{1}{24} \, q_{ijkl}^0 \, p_i \, p_j \, p_k \, p_l, \quad (1)$$

where $q$ is the vertical slowness $p_z$ as a function of the horizontal slowness $p_x$ and $p_y$. This way the use of the Christoffel equation $\Gamma(\mathbf{n})$ as a function of the wave normal vector n [2] and the Christoffel equation $\Gamma(\mathbf{p})$ as a function of the slowness vector p [7] resulted equivalent in the direct [10] and inverse [11] analysis of vertical seismic profiles in orthorhombic elastic media near the vertical axis of symmetry and with azimuthal angular dependence. The expression for the derivatives of the second order is given by the equation [7]:

$$q_{ij}^0 = \frac{\partial^2 q}{\partial p_i \partial p_j} = -\frac{F_{pi,pj} + F_{pi,q} \, q_j + F_{pj,q} \, q_i + F_{q,q} \, q_i \, q_j}{F_q}, \quad (2)$$

where $F$ as function of slowness is given by the determinant $F = Det\,[C_{ijlk} \, p_j p_l - \delta_{ik}] = 0$ and the general relation for the slowness vector in term of the phase elastic velocities $\mathbf{p} = (p_1, p_2, q) = V^{-1}(\theta_1, \theta_2)\,\mathbf{n}$. In this case the odd terms, $q_i^0 = q_{ijk}^0 = 0$ and the non-hyperbolic $q_{ijkl}^0$ moveout term is neglected in the hyperbolic analysis performed in this work.

In [17] using Eqs. 1 and 2, we were able to visualize the approximate solution for two azimuthal angles in orthorrombic media for the three modes using the slowness approach numerically given by Eq. 1 and the general solution of the Christoffel equation in terms of the normal vector for the Cracked Greenhorn shale [11]. Fig. 2 summarizes the results, where it can be seen that the transversal response impulse $S_1$ has a different azimuthal solution depending on Eq. 1 (blue ellipse) for the approximation, or for the general exact solution (red ellipse) as explained below. The other elastic modes (namely the P and $S_2$ modes) are well approximate in this case using the slowness approach (units for the elastic stiffnesses are given in CGS units [gr / (cm. s$^2$)] and the density $\rho = 1$ gr/cm$^3$

$$C_{ijlk} = \begin{pmatrix} 336.56 & 117.27 & 103.32 & 0 & 0 & 0 \\ 117.27 & 310.00 & 92.27 & 0 & 0 & 0 \\ 103.32 & 92.27 & 223.95 & 0 & 0 & 0 \\ 0 & 0 & 0 & 49.09 & 0 & 0 \\ 0 & 0 & 0 & 0 & 54.00 & 0 \\ 0 & 0 & 0 & 0 & 0 & 96.36 \end{pmatrix}$$

In particular, we used one general expression that condensed the three elastic fields, namely, the longitudinal (P) and the two transversal modes $S_1$ and $S_2$ (called sometimes SV and SH respectively) using the Voigt notation that changes the indices of the fourth order elastic tensor (stifnesses $c_{ijlk}$ or compliances $s_{ijlk}$) "ijlk" by the new ones "ij" one has [15,16]

$$\frac{1}{W_i(\phi_{1,i}, \phi_{2,i})} = \frac{1}{W_{i,z}} [\cos\phi_{1,i}]^2 + [\sin\phi_{1,i}]^2 \left\{ \frac{1}{W_{NMO,[xz]}^i} [\cos\phi_{2,i}]^2 + \frac{1}{W_{NMO,[yz]}^i} [\sin\phi_{2,i}]^2 \right\}, (3)$$

where the new parametrization in terms of the square velocities is given as follows: $W_i$ corresponds to the square of the group velocities times the density $\rho$ of the medium as a function of the group polar (first sub-index 1) azimuthal (second sub-index 2) angles $\phi_l$ according to Fig. 1 panel (a) and $W_i(\phi_{1,i}, \phi_{2,i}) = \rho\, v_i^2(\phi_{1,i}, \phi_{2,i})$ with the symbol i representing the three modes, i = P, $S_1$ or $S_2$. Figure 1 (where $P$ response impulses are shown in brown color and $Si$ modes in yellow color) shows the group polar and azimuthal angles and the group velocities (left a-panel) and the use in VSP profiles with "*P-Si*" conversion points at the subsurface interface with coordinates (x, y, z) (right b-panel)

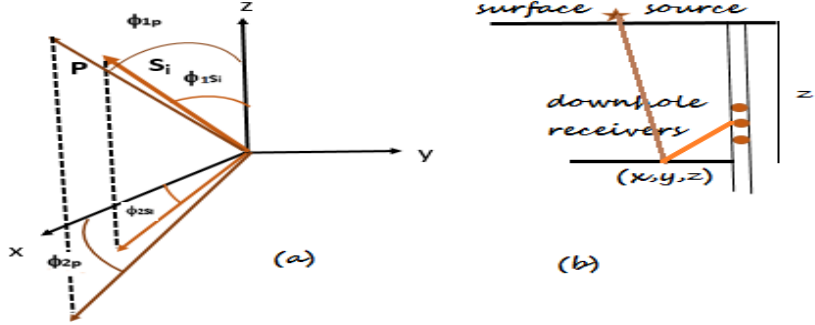

*Figure 1: Elastic response impulses with their respective azimuthal and polar angles in panel (a). Panel (b) corresponds to the vertical seismic profile with incident p wave and reflected $S_i$ waves.*

The normal move-out parametrization is specified by the coefficients in from of Eq. 3 that respectively are in the case of orthorhombic symmetry [15,16]:

$$W^P_{NMO,xz} = c_{55} + \frac{(c_{13} + c_{55})^2}{(c_{33} - c_{55})}, \quad W^P_{NMO,yz} = c_{44} + \frac{(c_{23} + c_{55})^2}{(c_{33} - c_{44})}, \quad (4)$$

$$W^{S1}_{NMO,xz} = c_{66}, \quad W^{S1}_{NMO,yz} = c_{22} + \frac{(c_{23} + c_{44})^2}{(c_{44} - c_{33})}, \quad (5)$$

and

$$W^{S2}_{NMO,xz} = c_{11} + \frac{(c_{13} + c_{55})^2}{(c_{55} - c_{33})}, \quad W^{S2}_{NMO,yz} = c_{66} \quad (6)$$

The results of azimuthal comparison for the three orthorhombic modes using Eqs. 4-6 and the exact solution given in [2] can be given by the expression $v_{gi} = {c_{ijkl}\, \alpha_j\, \alpha_k\, n_l}/{W_i^{1/2}}$ where the $W_i$ are the square of the phase velocities, the $c_{ijkl}$ are the elastic constants, the $\alpha$'s are the eigenvectors, and **n** is the normal direction, sketched in Fig. 1

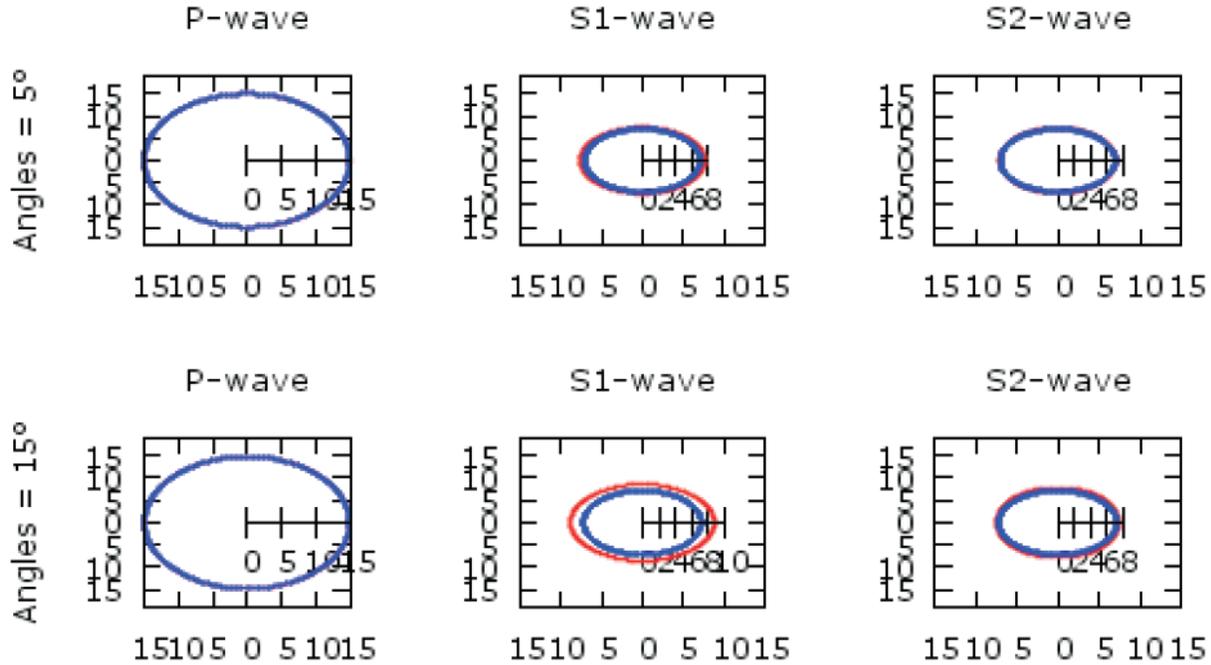

*Figure 2: Azimuthal visulization of the elastic response impulses in orthorrombic media according to [16,17] for two values of the polar in the case of the Cracked Greenhorn shale using the exact solution (red color) and the ellipsoidal approximation (blue color). The main difference is observed for the transversal elastic mode $S_1$ when the polar angle is 15 °.*

A similar analysis for 2D VTI elastic media analysis of elliptical dependence on the velocity and the elastic constant inversion procedure was realized in [18].

In general, one can solve the Christoffel equation using the eigenvalues of the phase velocities $\Gamma(\mathbf{n})$ matrix as a function of the wave front normal vector $\mathbf{n}$ for each vertical plane of symmetry, namely, the [XZ] and the [YZ] planes (the horizontal [XY] plane won't be considered in this case, but in [16]is given the procedure on how to solve for the [XY] case. We can Let us do it using the Cracked Greenhorn shale [11] separating the corresponding matrix for each vertical symmetry plane using the exact solution and the correspondent ellipsoidal approximation. The results are given in Figs 3 and 4.

Let us write the elastic matrix elements $c_{\alpha\beta}$ in Voigt notation. The orthorhombic group has nine independent elastic constants $c_{11}$, $c_{22}$, $c_{33}$, $c_{44}$, $c_{55}$, $c_{66}$, $c_{12}$, $c_{13}$, $c_{23}$ plus the three no diagonal depending ones given by the symmetric identity $c_{\alpha\beta} = c_{\beta\alpha}$, i.e., $c_{21}$, $c_{31}$, and $c_{32}$ making twelve in total [2]. It assembles a complicated numerical solution for group velocities that were visualized in [15,16] since we have to solve for both, the polar and azimuthal group angles.

In the case of a hexagonal (VTI medium) one has however five components for the [XZ] vertical plane, i.e., $c_{11}$, $c_{33}$, $c_{44}$, $c_{66}$, and the non-diagonal ones $c_{13}$, and the element $c_{12} = c_{11} - 2 c_{66}$. This case can be solving in a rather less complicated eigenvalues form than the orthorhombic one since one has a unique polar angle with respect to the vertical axis of symmetry [2]. The [YZ] vertical plane, is represented by the stifnesses $c_{22}$, $c_{33}$, $c_{55}$, $c_{66}$, and the non-diagonal ones $c_{23}$, and the element $c_{12} = c_{22} - 2 c_{66}$. All values are taken from the vertical symmetry planes od the Dellinger Cracked Greenhorn shale [11] with the orthorhombic set of fractures. In this work, we visualize independently for each vertical symmetry plane using an algebraic solver. The [XZ] and [YZ] solution for the group velocities with the Greenhorn shale partial matrices is represented in Fig 3 for both symmetry planes.

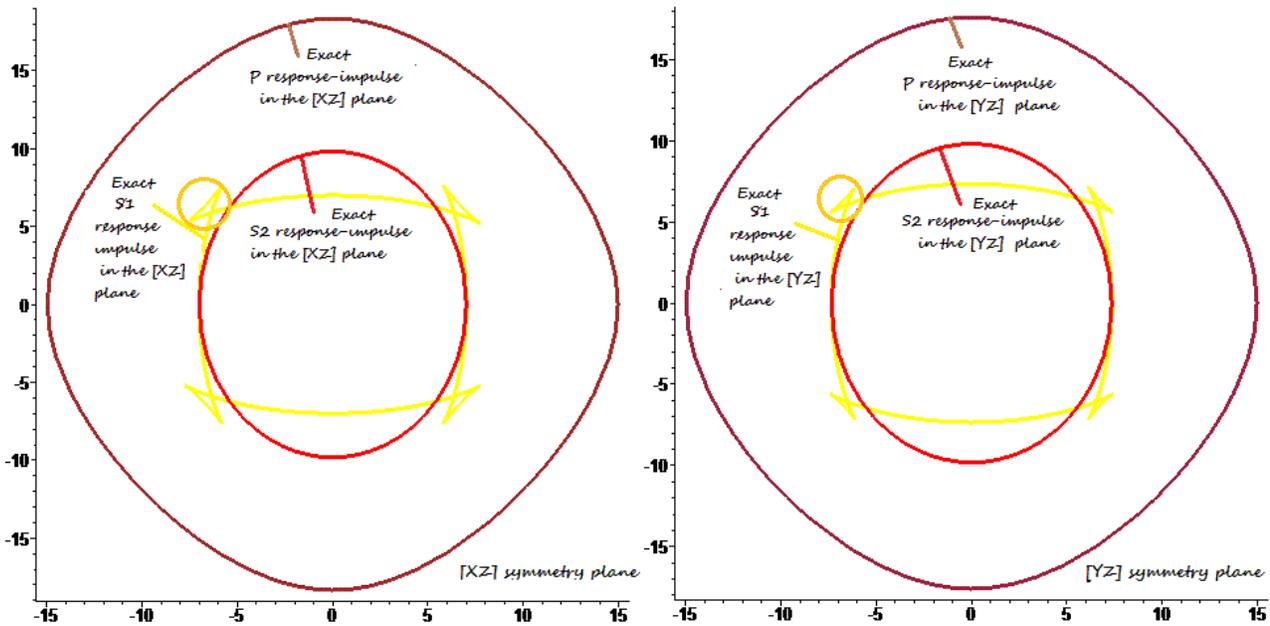

*Figure 3 : Exact solution of the Greenhorn shale case in the vertical planes of symmetry for a homogeneous orthorhombic anisotropy.*

The main difference for the Greenhorn shale simulation of the three exact elastic group waves according to Fig. 3 is seen mainly in the $S_2$ component where the triplication (defined here as three different values of the $S_1$ group velocity with a fixed polar group angle around 45 degrees) are more noticeable in the [XZ] than in the [YZ] symmetry plane (see both orange circles in the right and left panes of Fig. 3. This qualitative is explained because in the panel a, the elastic constant that controls the triplication has a value of $c_{13}$ = 103.3 gr / (cm. $s^2$) meanwhile the stiffness that control the $S_1$ group velocity in the [YZ] plane $c_{23}$ = 93.3 gr / (cm. $s^2$), i.e. $c_{13} \geq c_{23}$ .

In the case of the ellipsoidal approximation of [15,16] the same calculation with Eqs. 4-6 gives are presented in Fig. 4. It can be seen that for the case of the right panel in Fig. 4 representing the ellipsoidal case of the [XZ] symmetry plane and where the elastic constant $c_{13}$ is bigger and gives an augmented triplication conducting to a higher relative error in velocity values with values smaller than the $S_2$ mode, meanwhile in the right panel of Fig. 4, the shear vertical velocity $S_1$ has a higher set of values than the $S_2$ horizontal shear mode.

In the Fig. 8 of reference [16] it was observed a constant and higher relative error in the inversion of $c_{13}$ for the whole range of azimuthal group angles in the [XZ] symmetry vertical plane. In the [YZ] plane of symmetry, where the elastic constant input value $c_{23}$ is smaller in velocity calculations, it resulted in smaller relative errors in the inversion calculation of $c_{23}$ relaying on certain accuracy (right panel of Fig. 4) in the correct positioning of the shear vertical $S_1$ and shear horizontal $S_2$ modes. Finally, this is also seen in Fig. 9 of reference [16] when the relative error in the inversion of $c_{23}$ is smaller and grows slightly with azimuth than the case for $c_{13}$ for which the relative inversion error is always constant, clarifying the differences in the ellipsoidal inversion of $c_{13}$ and $c_{23}$ [16].

Based on the previous analysis one can see why the ellipsoidal approximation works well in case of the cracked Greenhorn shale [11] with small differences in the elastic stiffnesses $c_{13}$ and $c_{23}$ and it can finally define the ellipsoidal orthorrombic approximation as the one that works well in VSP geometries with small practical differences in elastic stifnesses values for $c_{13}$ and $c_{23}$ and where the vertical symmetry planes are set close in values to each other for a cracked set of perpendicular fractures in a homogeneous elastic medium.

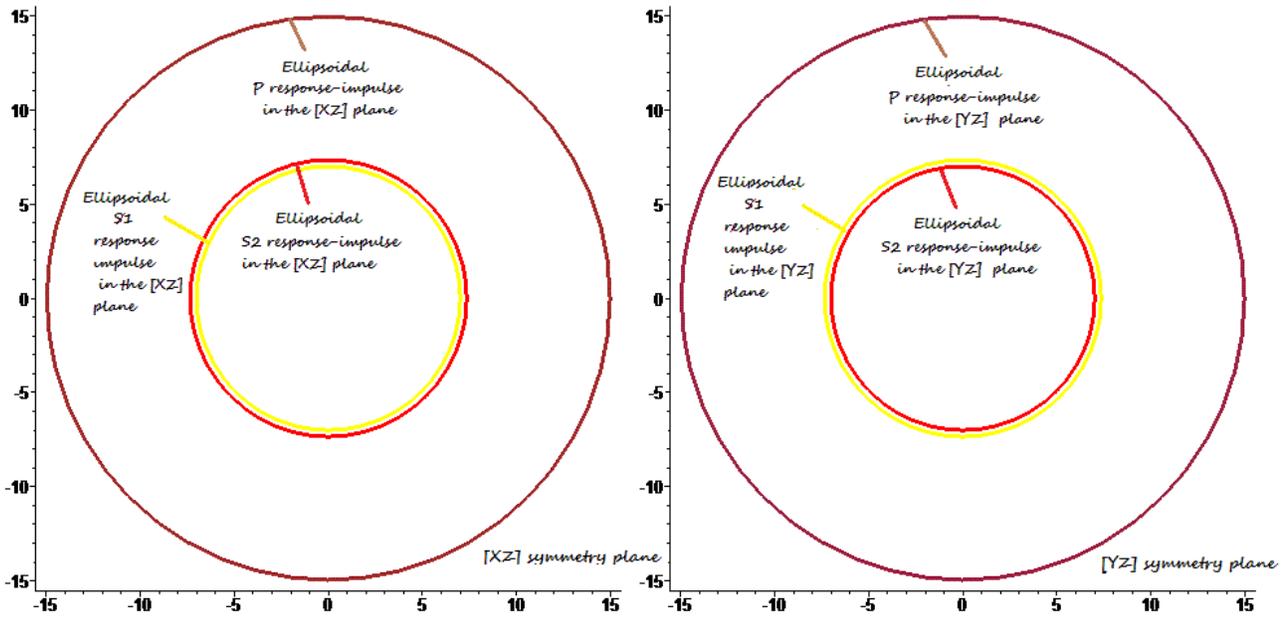

*Figure 4: Ellipsoidal aproximation using the same elastic stifnesses of Fig. 3. Notice the difference in the group velocities $S_1$ and $S_2$ where the group velocity $S_1$ is smaller than $S_2$ if the stiffness $C_{13}$ has a higher value in the [XZ] plane (left panel) than $C_{23}$ in the [YZ] plane.*

**KINEMATICAL DERIVATION OF THE CONVERSION POINT IN ORTHORROMBIC MEDIA NEAR THE VERTICAL AXIS OF SYMMETRY.**

On the other hand, seismic velocity analysis represents one of the most widely used processing techniques for the treatment of geophysical seismic data in the oil and gas venture. A method of acquisition commonly used is the Common-Mid-Point (CMP) method. The general idea of this method is to acquire a series of traces (gather) that reflect from the same common subsurface mid-point. Similarly, to group seismic traces of multicomponent signals under the CMP criterion. Thus, an expression for the conversion point of each of the traces can be used to group those that have the same CMP. General expressions were developed previously for the calculation of the conversion of "$P - Si$" points in isotropic [19], vertical transverse isotropy (VTI) [20], horizontal transversal isotropy (HTI) [21], and orthorhombic media [22] with the goal of seismic exploration analysis.

However, these expressions can be simplified when it comes to performing analysis in a medium with azimuthal anisotropy such as fractured orthorhombic media with the dependency on several elastic stiffnesses. In general, azimuthally elastic depend on media such as the HTI, the orthorhombic, and the monoclinic ones, are those with combinations of multiple vertical fracture sets and possible horizontally fine layering, they are of great importance for fracture characterization [4].

Therefore, it is convenient to obtain expressions for the coordinates of the conversion "P - Si" point considering azimuthal variations of the velocity for any mode of propagation for specific acquisition geometries such as VSP. In section it will be shown, how to apply a particular methodology developed for VTI media [20] to derive asymptotic 3D analytic expressions for the conversion point in orthorhombic media within the ellipsoidal approximation [15,16] for the P-S1 and P-S2 modes. We should note that such expressions will be valid near a single axis of symmetry, such that, those with large "depths" and small "offsets", represents practical cases encountered in seismic velocity analysis for vertical seismic profiles and where the z location of the surface source and the downhole receiver is known.

Let us consider the case of a general homogeneous elastic medium with orthorhombic symmetry and a particular density that for simplicity we fixed equal to 1 gr/cm³, as represented in Fig. 5 where the two vertical planes [XZ] and [YZ] and the ellipsoidal group velocities will be combined with the help of trigonometric analysis, the use of the Fermat theorem and the derivation of travel times following [20], arriving to analytical expressions of the conversion point in the ellipsoidal anisotropic case.

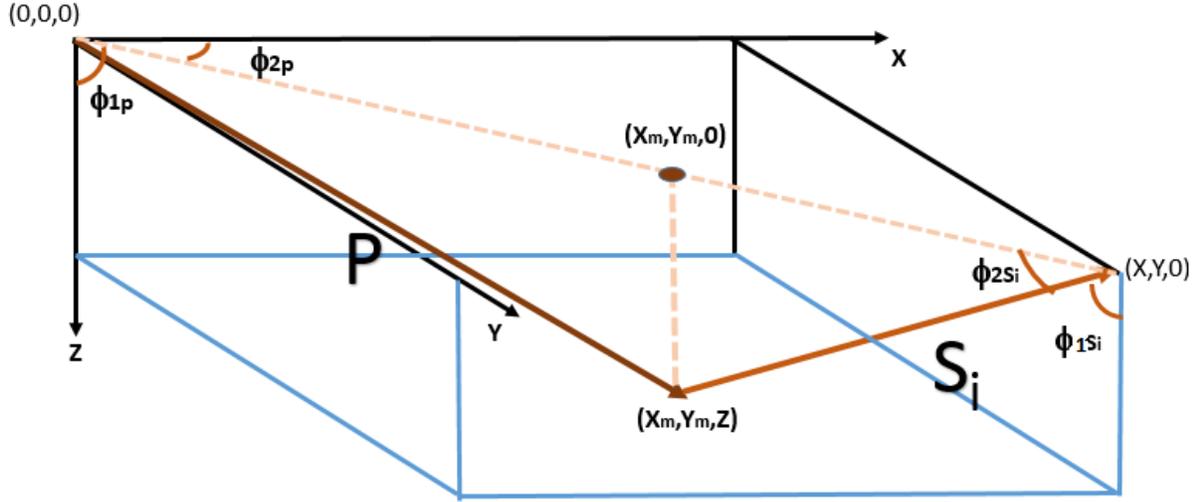

*Figure 5: General scheme of the P-$S_i$ conversion point on a 3-D line in a homogeneous anisotropic medium with orthorhombic symmetry.*

From Fig. 5, it can be seen that for the incident P response-impulse, the following equations are true as function of the polar (denoted by index 1) and azimuthal group angles (denoted by index 2):

$$\cos^2\phi_{1P} = \frac{Z^2}{Z^2 + X_j^2 + Y_j^2}, \quad \sin^2\phi_{1P} = \frac{X_j^2 + Y_j^2}{Z^2 + X_j^2 + Y_j^2}, \quad \cos^2\phi_{2P} = \frac{X_j^2}{X_j^2 + Y_j^2}, \quad \sin^2\phi_{2P} = \frac{Y_j^2}{X_j^2 + Y_j^2} \quad (7)$$

and for the reflected shear response-impulses ($S_i$ with the index i = 1 = SV, and or 2 = SH). It follows that as a function of the shear correspondent azimuthal angles that control the wave-front in the horizontal plane [XY] one has

$$\cos^2\phi_{1Sj} = \frac{Z^2}{Z^2 + (X-X_j)^2 + (Y-Y_j)^2}, \quad \sin^2\phi_{1Sj} = \frac{(X-X_j)^2 + (Y-Y_j)^2}{Z^2 + (X-X_j)^2 + (Y-Y_j)^2} \quad (8)$$

By the geometrical congruence of similar triangles it follows also that $\phi_{2P} = \phi_{2Sj}$, and therefore the following relationship takes place:

$$\cos^2\phi_{2Sj} = \cos^2\phi_{2P} = \frac{X_j^2}{X_j^2 + Y_j^2} \quad \text{and} \quad \sin^2\phi_{2Sj} = \sin^2\phi_{2P} = \frac{Y_j^2}{X_j^2 + Y_j^2} \quad (9)$$

For equations (8) and (9) $j = 1,2$ are the two propagation modes $S_1$ y $S_2$ respectively; $\phi_{1Sj}$ and $\phi_{2Sj}$ are the polar and azimuthal group angles for the propagation modes $S_j$;

The kinematic travel time analysis procedure is used a continuation with the aim of the Fermat theorem [20]. The expression that determines the travel-time of the incident $P$ wave and reflected $S_i$ waves is the sum of the two travel-times $T = T_{Pi} + T_{Sj}$ where each travel time corresponds to the expressions to be derived using the notation from the previous section and Fig. 3 where $T_{Pi}$ is the travel time from the incident $P_i$ wave and $T_{Sj}$ is the travel time for the reflected $S_j$ wave. Considering elementary trigonometry and looking at Fig. 3, an expression for both travel times, i.e., $T_{Pi}$ y $T_{Sj}$ as function of the group angles and the elastic constants can be conveniently written as $T_{Pi} = \frac{Z}{V_{P(\emptyset_{1P},\emptyset_{2P})} \cos \emptyset_{1P}}, T_{Sj} = \frac{Z}{V_{Sj(\emptyset_{1Sj},\emptyset_{2Sj})} \cos \emptyset_{1sj}}$ where $v_P$ and $v_{Sj}$ are the group velocities for the longitudinal and transversal elastic modes, respectively, where by means of some straightforward but lengthy algebra a more suitable for the analysis in orthorrombic fracture media near the vertical axis, parametrized expressions as functions of the square of the normal move-out velocities (defined as the hyperbolic velocities for the travel time with small-spread and a small-dip approximation) for the converted $P$ - $S_i$ modes are derived.

The square of the response-impulses as function of the azimuthal angles can be given by the mathematical expresion $W_{gi}^{-1}(\emptyset_{1,i},\emptyset_{2,i}) = C_0^i + C_1''^i(\emptyset_{2,i})\sin^2\emptyset_{1,i}$ where $C_0^i$ is related to the velocity of propagation in the vertical axis, and $C_1''^i$ has the following anisotropic dependence form in the ellipsoidal case $C_1''^i(\emptyset_{2,i}) = W_{NMO[x,z]}^{-1,i} - W_z^{-1,i} + (W_{i,NMO[y,z]}^{-1} - W_{i,NMO[x,z]}^{-1})\sin^2\emptyset_{2,i}$.

Taking into account the mathematical expression for the different propagation modes developed in the previous paragraphs and considering that the group polar angles are small but the azimuthal angles can have any allowed value in the horizontal plane, the travel times expressions for longitudinal and transversal group velocities in ellipsoidal orthorhombic elastic media can be written using the geometry considerations sketched in Fig. 5.

Henceforth, for the P wave the incident 3D azimuthal travel time takes the form

$$T_{Pi} = \frac{1}{4\sqrt{C_0^P}Z}(2Z^2 + X_1^2 + Y_1^2)(2C_0^P + C_1''^P \sin^2(\emptyset_{1,p}), \quad (10)$$

and also for the S₁ and S₂ transversal waves, the reflected travel times are given accordingly to

$$T_{Sj} = \frac{1}{4\sqrt{C_0^S}Z}(2Z^2 + (X - X_1)^2 + (Y - Y_1)^2)(2C_0^S + C_1''^S \sin^2(\emptyset_{1,s}), \quad (11)$$

Once an approximate expression of the ray paths travel times of the different response impulses have been obtained, the conversion point can be found, using Fermat's theorem of the minimum time considering that in general one has *(X_j, Y_j) < (X, Y)*.

In the ellipsoidal orthorrombic elastic case, we start from the Fermat theorem for each horizontal azimuthal component $\frac{\partial T}{\partial X_j} = \frac{\partial T}{\partial Y_j} = 0$. On the other hand, for the Z direction, the velocity is constant, we suppose that the depth Z is known (particularly this is a fact in the case of vertical seismic profiles), and therefore, asymptotic analysis is performed only for the Xj and Yj components. Taking into account the total horizontal travel time for the three responses impulses and their respectively derivatives using the partial derivatives expression $\frac{\partial T}{\partial X_j} = \frac{\partial T_{Pi}}{\partial X_j} + \frac{\partial T_{Sj}}{\partial X_j}$, considering Eq. (10) the following general expresion near the vertical axis is derived after some algebraic manipulations $\frac{\partial T}{\partial X_j} = \frac{1}{Z} \left\{ \frac{X_j}{\sqrt{C_0^P}} W_{NMO[x,z]}^{-1,P} + \frac{X_j - X}{\sqrt{C_0^{Sj}}} W_{NMO[x,z]}^{-1,Sj} \right\} = 0$, which means that $\left\{ \frac{X_j}{\sqrt{C_0^P}} W_{NMO[x,z]}^{-1,P} + \frac{X_j - X}{\sqrt{C_0^{Sj}}} W_{NMO[x,z]}^{-1,Sj} \right\} = 0$.

Solving for $X_j$ one arrives at the first general expression of one of the conversion point components for orthorrombic ellipsoidal media near the vertical axis of symmetry as function of normal moveout square velocites

$$X_j = X \frac{\sqrt{C_0^P} W_{NMO[x,z]}^{-1,Sj}}{\sqrt{C_0^P} W_{NMO[x,z]}^{-1,Sj} + \sqrt{C_0^{Sj}} W_{P,NMO[x,z]}^{-1}} . \quad (12)$$

Similarly using $\frac{\partial T}{\partial Y_j} = \frac{\partial T_{Pi}}{\partial Y_j} + \frac{\partial T_{Sj}}{\partial Y_j}$, multiplying the traveltime derivative by $Z\sqrt{C_0^P C_0^{Sj}}$ and making the respective factorization, a general expression can be found for the second horizontal component $Y_j$ considering Eq. (11) to get

$$Y_j = Y \frac{\sqrt{C_0^P} W_{NMO[y,z]}^{-1,Sj}}{\sqrt{C_0^P} W_{NMO[y,z]}^{-1,Sj} + \sqrt{C_0^{Sj}} W_{NMO[y,z]}^{-1,P}} . \quad (13)$$

In this way, it has been possible to obtain asymptotic expressions for the coordinates of the conversion point for a medium with orthorhombic elliptical anisotropy considering azimuthal group angle variations that represents a perpendicular set of fractures.

The specific form of the horizontal coordinates in the [XY] plane for the conversion point in an orthorhombic medium for the conversion form is for the $P - S_1$ conversion point is the $(X_1, Y_1)$ pair

$$(X_1, Y_1) = \left( X \frac{\sqrt{C_0^P} W_{NMO[x,z]}^{-1,S1}}{\sqrt{C_0^P} W_{NMO[x,z]}^{-1,S1} + \sqrt{C_0^{S_1}} W_{NMO[x,z]}^{-1,P}}, Y \frac{\sqrt{C_0^P} W_{NMO[y,z]}^{-1,S1}}{\sqrt{C_0^P} W_{,NMO[y,z]}^{-1,S1} + \sqrt{C_0^{S_1}} W_{NMO[y,z]}^{-1,P}} \right) \quad (14)$$

For the $P - S_2$ conversion point analously we find the $(X_2, Y_2)$ pair

$$(X_2, Y_2) = \left( X \frac{\sqrt{C_0^P} W_{NMO[x,z]}^{-1,S2}}{\sqrt{C_0^P} W_{NMO[x,z]}^{-1,S2} + \sqrt{C_0^{S_2}} W_{NMO[x,z]}^{-1,P}}, Y \frac{\sqrt{C_0^P} W_{NMO[y,z]}^{-1,S2}}{\sqrt{C_0^P} W_{NMO[y,z]}^{-1,S2} + \sqrt{C_0^{S_2}} W_{NMO[y,z]}^{-1,P}} \right) \quad (15)$$

This is important for different reasons such as the analysis of seismic data acquired in multicomponent VSP surveys, which is crucial for the correct interpretation of the data (positioning of events and imaging). Physically in the interface where the conversion occurs, the energy carried by the longitudinal response-impulse is converted into energy carried now by the transversal modes.

One way to corroborate the general validity of these results is the reduction from Eqs. (14) and (15) to the isotropic limit [19]. For the isotropic case, we have that for the longitudinal response impulses it follows that the new normal move-out square velocities are $W_{P,NMO[x,z]}^{-1} = W_{P,NMO[y,z]}^{-1} = C_0^P = \frac{1}{V_P^2}$, and for the transversal elastic modes one has that the square of the normal move-out velocities are bring down to $W_{Sj,NMO[x,z]}^{-1} = W_{Sj,NMO[y,z]}^{-1} = C_0^{Sj} = \frac{1}{V_S^2}$. Replacing those expressions back in Eqs. (14) and (15) one finally obtains $(X_j, Y_j) = (X \frac{1}{1+\frac{V_S}{V_P}}, Y \frac{1}{1+\frac{V_S}{V_P}})$ which are the well-known expression for an isotropic elastic medium [19].

## CONCLUSIONS AND RECOMMENDATIONS

First, we conclude that simple analytical expressions using the Fermat theorem and for practical multi-walkaway velocity analysis such as in vertical seismic profiles of the conversion "P - Si" points that belong to a single interface in orthorhombic ellipsoidal elastic media with general azimuthal dependence (including near offsets and small polar angles) were derived and compared with the isotropic case.

Second, we explain the difference in the inversion of the elastic constants $c_{13}$ and $c_{23}$ within the ellipsoidal approximation [16] with respect to the differences in the growth of the azimuthal angle. The relative error in the inversion of $c_{13}$ remained constant for azimuthal angle values in between 0° and 90°, meanwhile, the relative errors in the inversion of the stiffness $c_{23}$ increased with the growth of the azimuthal angle (see Figs. 7 and 8 in reference [16] respectively). We find that is due to the size of the triplication in each vertical symmetry plane and as it can be seen from Fig. 3, higher stiffness input values for $c_{13}$ set constant inverted values for arbitrary azimuthal angles and constant higher relative errors.

As additional points, a very brief analysis of the use of the Christoffel equation in different representations with the respective references is mentioned for future works. Finally, we recommend the study of orthorhombic systems using the same methodology as the one presented in [23], in particular, to observe the P response impulse with imaging purposes implementing perfect matched layer boundary conditions, staggered finite-difference grid 3D schemes, and Valgrind's memcheck, to simulate a 3D elastic wave generating synthetic seismograms and screenshots. We also point out the inclusion of pressure for studies and visualization in anisotropic media using VSP profiles with downhole receivers as the one presented in the work [24].

## ACKNOWLEDGEMENTS

P. Contreras acknowledges the support and research discussions in Intevep SA, the Research branch of Petroleos de Venezuela from the years 1995 to 1998.

## REFERENCES

[1] Grechka, V. 2017. Algebraic degree of a general group-velocity surface. Geophysics 82: WA45-WA53. DOI: https://doi.org/10.1190/geo2016-0523.1

[2] M. J. Musgrave, 1970. Crystal Acoustics. Holden Day.

[3] L. Thomsen 1986. Weak elastic anisotropy. Geophysics 51: 1954-1966 DOI: https://doi.org/10.1190/1.1442051

[4] Tsvankin, I., and V. Grechka, 2011. Seismology of azimuthally anisotropic media and seismic fracture characterization: Society of Exploration Geophysicists DOI: https://doi.org/10.1190/1.9781560802839

[5] Pedro Contreras, Vladimir Grechka, and Ilya Tsvankin 1999 Moveout inversion of *P*-wave data for horizontal transverse isotropy Geophysics 64:4, 1219-1229 DOI: https://doi.org/10.1190/1.1444628

[6] Vladimir Grechka and Ilya Tsvankin, (1997), "Moveout velocity analysis and parameter estimation for orthorhombic media," *SEG Technical Program Expanded Abstracts*: 1226-1229. DOI: https://doi.org/10.1190/1.1885619

[7] Qi Hao, Alexey Stovas 2014. Anelliptic approximation for P-wave phase-velocity in orthorhombic media. 16$^{th}$ International Workshop on Seismic Anisotropy (16IWSA), Natal, Brazil. DOI: 10.13140/2.1.3970.8488

[8] Vladimir Grechka, Pedro Contreras, and Ilya Tsvankin Inversion of normal moveout for monoclinic media SEG Technical Program Expanded Abstracts 1999. January 1999, 1883-1886 DOI: https://doi.org/10.1190/1.1820913

[9] Grechka, Vladimir. 2020. Christoffel equation in polarization variables. Geophysics 85(3):C91. DOI: https://doi.org/10.1190/geo2019-0514.1

[10] Alexey Stovas. 1998. Wave-equation-based normal moveout 1998. Journal of Seismic Exploration 7(1):1-8

[11] Levin, F. K., 1978. The reflection, refraction and diffraction of waves in media with elliptical velocity dependence, Geophysics, 43(3):528-537. DOI: https://doi.org/10.1190/1.1440833

[12] Byum, S., 1982. Seismic parameters for media with elliptical velocities dependencies. Geophysics 47:1621-1626. DOI: https://doi.org/10.1190/1.1441312

[13] Muir, F., 1990. Various equations for TI media. SEP Stanford 70:367-372

[14] J. Dellinger 1991. Anisotropic seismic-wave propagation Ph.D. Thesis Stanford University.

[15] Contreras, Pedro, Klíe, Héctor, Mora, Carmen and Michelena, Reinaldo Ellipsoidal Approximation of Velocities in Orthorhombic Media, European Association of Geoscientists & Engineers. Conference Proceedings, 5th International Congress of the Brazilian Geophysical Society, Nov 1997, cp-299-00087 DOI: https://doi.org/10.3997/2214-4609-pdb.299.87

[16] Pedro Contreras, Héctor Klíe, and Reinaldo J. Michelena Estimation of elastic constants from ellipsoidal velocities in orthorhombic media SEG Technical Program Expanded Abstracts 1998. January 1998, 1491-1494 DOI: https://doi.org/10.1190/1.1820194

[17] Pedro Contreras, Luis Rincon, Jose Burgos Ellipsoidal response-impulses in fracture media 2014 Conference: 12th International Congress on Numerical Methods in Engineering and Applied Sciences · CIMENICS 2014 Volume: I, 25- 30 DOI: https://doi.org/10.48550/arXiv.1812.09125

[18] R. Michelena 1994. Elastic constants of transversely isotropic media from constrained aperture travel times. Geophysics 59: 658-667 DOI: https://doi.org/10.1190/1.1443625

[19] Fromm et al. 1985. Static and dynamic correction, in Dohr, G. Ed., Seismic shear waves. Handbook of geophysical exploration, Geophysical Press, 15b:191-225.

[20] Sena, A. and Toksöz, N. 1993, Kirchhoff migration and velocity analysis for converted and no converted waves in anisotropic media," Geophysics, 58:265-275. DOI: https://doi.org/10.1190/1.1443411

[21] Qi Hao, Alexey Stovas, and Tariq Alkhalifah, (2013), "The azimuth-dependent offset-midpoint travel time pyramid in 3D HTI media," SEG Technical Program Expanded Abstracts: 3335-3339 DOI: https://doi.org/10.1190/segam2013-0058.1

[22] Shibo X., and A. Stovas, 2019. Estimation of the conversion point position in elastic orthorhombic media, Geophysics 84:C15-C25. DOI: https://doi.org/10.1190/geo2018-0375.1

[23] P. Contreras, Larrazábal, and C. Florio, 2019. Perfectly matched layers' simulation in two-dimensional media. Canadian Journal of Pure and Applied Sciences Vol. 13, No. 3, pp. 4861-4867. Online ISSN: 1920-3853; Print ISSN: 1715-9997

[24] Guerrero, Oscar. (2022). My notes on the prediction of pressure induce phonon instabilities via computational modeling of Acoustic wave propagation in isotropic and anisotropic crystals. DOI: 10.13140/RG.2.2.34112.81925